\documentclass{iopart}
\usepackage{graphicx}
\usepackage{bm}

\newcommand{\hG}{\hat{G}}
\newcommand{\hGG}{\hat{\cal G}}
\newcommand{\hH}{\hat{H}}
\newcommand{\bZ}{\bar{\cal Z}}
\newcommand{\bg}{\bar{g}}

\newcommand{\cZ}{{\cal Z}}

\begin{document}
\title[Green's functions of electrons in (Mo,W)X$_2$ in a magnetic field]
      {Green's functions of electrons in group-VI dichalcogenides in a magnetic field}
\date{\today}
\author{Tomasz M Rusin$^1$ and Wlodek Zawadzki$^2$}
\address{ $^1$ Orange Polska sp. z o. o., Al. Jerozolimskie 160, 02-326 Warsaw, Poland\\
          $^2$ Institute of Physics, Polish Academy of Sciences, Al Lotnik\'ow 32/46, 02-668 Warsaw, Poland}
 \ead{Tomasz.Rusin@orange.com}

\submitto{\JPA}

\begin{abstract}
Closed expression for the Green's function of the stationary two-dimensional
Schrodinger equation for an electron in group-VI dichalcogenides in the presence of a
magnetic field is obtained in terms of the Whittaker functions. The resulting Green's function
operator is a~$8 \times 8$ matrix consisting of four block-diagonal~$2\times 2$ matrices,
each of them characterized by different values of valley index and electron spin.
The obtained results are used to calculate
local density of states induced by a neutral delta-like impurity in the presence
of a magnetic field within the lowest Born approximation. The spatial electron density
experiences either the Friedel-like oscillations or an exponential decay as a function
of the distance from the impurity.
\end{abstract}

\maketitle

\section{Introduction}

Some years ago Dodonov et al~\cite{Dodonov1975} calculated the stationary Green's
function of a free 2D electron in a homogeneous magnetic field and obtained analytical
results in terms of the Whittaker functions. Similar problems were recently
investigated for low-dimensional systems~\cite{Gusynin1995,Gorbar2002,Murguia2010},
where Green's functions were obtained as infinite sums of Laguerre polynomials.
Horing and Liu~\cite{Horing2009} obtained a propagator as an infinite sum and, alternatively, as the
second solution of the Bessel wave equation. A closed form of the propagator in monolayer
graphene in terms of the confluent hypergeometric function was obtained by Piatkovskii and
Gusynin~\cite{Pyatkovskii2011} and Gamayun et al~\cite{Gamayun2011}. The present authors~\cite{Rusin2011}
calculated the propagator in monolayer and bilayer graphene and obtained results
in terms of Whittaker functions. Ardenghi et at~\cite{Ardenghi2015} calculated
the Green's function of graphene taking into account corrections caused by the
coherent potential approximation, and Gutierrez-Rubio at al~\cite{Rubio2016}
calculated numerically magnetic susceptibility of graphene and MoS$_2$
within the Green's function formalism.

Recently there has been increased interest in the synthesis as
well as in experimental and theoretical properties of
materials having two-dimensional hexagonal lattices as, e.g. MoS$_2$,
and other group-VI dichalcogenides.
There are many similarities between group-VI dichalcogenides and monolayer graphene,
but there are two significant differences. First, the presence of
two different atoms in the crystal lattice in group-VI dichalcogenides
leads to the energy gap~$1.0-1.6$~eV in the electron band structure.
Second, in group-VI dichalogenides there exists strong spin-orbit coupling,
originating from the metal~$d$-orbitals. As a result, the conduction band-edge
state remains spin degenerate at~${\bm K}$ and~${\bm K'}$ points of the Brillouin zone~(BZ),
whereas the valence-band-edge state splits into two subbands separated by~$2\lambda \in (0.1-0.5)$~eV.

The energy band structure of group-VI dichalogenides is well parameterized
by the six-band tight-binding Hamiltonian~\cite{Liu2013}. However, a simpler quasi two band model
can be obtained within the~${\bm k} \cdot {\bm p}$ theory~\cite{Xiao2012}.
The resulting~$8 \times 8$ Hamiltonian consists of four~$2\times 2$ blocks corresponding to different
combinations of the electron spin and the valley indexes. After symmetrization and
adjustment of the energy scales, each~$2\times 2$ block of the~${\bm k} \cdot {\bm p}$
Hamiltonian is analogous to the~$2\times 2$
Hamiltonian of electrons in monolayer graphene with a nonzero energy gap.

In the present paper we calculate the electron Green's function operator for group-VI dichalogenides
in the presence of a magnetic field. The Green's function is a~$8 \times 8$ operator
consisting of four block-diagonal~$2\times 2$ operators.
The elements of the Green's function matrix are calculated in analogy
to that outlined in~\cite{Rusin2011} for gapless monolayer graphene,
For group-VI dichalogenides these elements are also expressed in terms
of the Whittaker functions. The present calculations are extensions of
the results in~\cite{Rusin2011} for nonzero energy gap and presence of the spin-orbit interactions.
To our knowledge, the expression
for Green's function for group-VI dichalogenides in terms of the Whittaker
functions has not been published.
The Green's function is applied to a calculation of the electron
density induced by a delta-like potential. As a result, the Friedel-like oscillations
in group-VI dichalogenides in a magnetic field are obtained. 
Early references to the Friedel oscillations in 3D electron gas in a
magnetic field are~\cite{Rensink1968,Glasser1969}. 
The Friedel oscillations in 2D electron gas
in a magnetic field were briefly treated by Simion and Giuliani~\cite{Simion2005} for a
simple parabolic energy band.

\section{Green's function}

In the quasi two band~${\bm k} \cdot {\bm p}$ theory the Hamiltonian for
group-VI dichalcogenides materials in the vicinity of~${\bm K}$ and~${\bm K'}$ points
of the BZ is~\cite{Xiao2012}
\begin{equation} \label{hH0}
 \hH = a_l t(\tau\hat{\sigma}_x k_x + \hat{\sigma}_y k_y) + \frac{\Delta}{2}\hat{\sigma}_z
      -\lambda \tau \frac{\hat{\sigma}_z-1}{2} \hat{s_z},
\end{equation}
where ${\bm k}$ is the wave vector,~$\tau=\pm 1$ is the
valley index,~$\hat{\bm \sigma}$ are the Pauli matrices,~$a$ is
the lattice constant,~$t$ is the effective hopping integral,~$\Delta$ is the
energy gap,~$\lambda$ is spin splitting at the valence band edge caused by
the spin-orbit interaction, and~$\hat{s}_z$ is the Pauli matrix for electron spin.
The Hamiltonian~(\ref{hH0}) describes the~$8 \times 8$ operator
with four uncoupled~$2\times 2$ blocks, and each block is characterized
by a different combination of~$\tau$ and~$s_z$.
The material parameters entering~(\ref{hH0}) are listed in~\cite{Xiao2012} and quoted in Table~1.

It is convenient to rewrite the Hamiltonian~(\ref{hH0}) in a more symmetric form
\begin{equation} \label{hH1}
 \hH = a_l t(\tau\hat{\sigma}_x k_x + \hat{\sigma}_y k_y) +
           g \hat{\sigma}_z + E_{s_z},
\end{equation}
in which~$g=\Delta/2 - E_{s_z}$ is half of the energy gap
and, for given~$s_z=\pm 1$ and~$\tau$, the energy~$E_{s_z}=s_z\tau\lambda/2$
shifts the zero of energy scale.
The symmetric form of the Hamiltonian allows us to apply a similar
methodology for Green's function calculation as that used for gapless and
gapped graphene~\cite{Pyatkovskii2011, Gamayun2011, Rusin2011}.
Energies~$g$ and~$E_{s_z}$ take two values:~$g =\Delta/2 \pm \lambda/2$
and~$E_{s_z}=\pm \lambda/2$, respectively. For all materials listed in Table~1
there is:~$\Delta \gg \lambda$, so that the energy gaps for both spin orientations
are close to each other. The eigenenergies of the Hamiltonian in (\ref{hH1}) are
\begin{equation} \label{eH1}
 E_{ks_z} = \pm \sqrt{(a_ltk)^2 + g^2} + E_{s_z}.
\end{equation}

\begin{table}
\label{Table_1}
\caption{Parameters used in Hamiltonian~(\ref{hH0}) for four group-VI dichalcogenides, after~\cite{Xiao2012}.}
\begin{tabular}{|c|c|c|c|c|}
\hline
Material &~$a_l$~(\AA) &~$t$~(eV) &~$\Delta$~(eV) &~$2\lambda$~(eV) \\
 \hline
  MoS$_2$ & 3.193 & 1.66 & 1.10 & 0.15 \\
  WS$_2$  & 3.197 & 1.79 & 1.37 & 0.43 \\
  MoSe$_2$& 3.313 & 1.47 & 0.94 & 0.18 \\
  WSe$_2$ & 3.310 & 1.60 & 1.10 & 0.46 \\
 \hline
\end{tabular}
\end{table}

In the presence of a magnetic field the Hamiltonian (\ref{hH1}) reads
\begin{equation} \label{hH2}
 \hH = \frac{a_l t}{\hbar}(\tau\hat{\sigma}_x \hat{\pi}_x + \hat{\sigma}_y \hat{\pi}_y) +
           g \hat{\sigma}_z + E_{s_z},
\end{equation}
where~$\hat{\bm \pi} = \hat{\bm p} + e{\bm A}$,~$\hat{\bm p}$ is
the electron's momentum,~${\bm A}$ is the vector potential, and the charge~$e>0$.
In the Landau gauge:~${\bm A}=(-By, 0)$. Let~$L=\sqrt{\hbar/eB}$ be the
magnetic radius and~$\xi =y /L-k_xL$.
Defining the standard raising and lowering operators for the
harmonic oscillator:~$\hat{a}ˆ=(\xi +\partial /\partial \xi)/\sqrt{2}$
and~$\hat{a}^+=(\xi -\partial /\partial \xi)/\sqrt{2}$,
the Hamiltonian for~$\tau=+1$ becomes
\begin{equation} \label{hH3}
\hH = \left(\begin{array}{cc} g & -\hbar \Omega \hat{a} \\
 -\hbar \Omega \hat{a}^+ & -g \end{array}\right) + E_{s_z},
\end{equation}
where~$\Omega =\sqrt{2} a_lt/(\hbar L)$. The eigen-energies and eigen-states of~$\hH$ are
\begin{eqnarray} \label{Enkx}
 E_{nk_x \eta s_z} &=& \eta \sqrt{n \hbar^2 \Omega^2 + g^2} + E_{s_z} \equiv \eta E + E_{s_z} \\
 \label{phiPls}
 \psi_{nk_x+}({\bm \rho}) &=& \frac{e^{ik_xx}}{\sqrt{\pi(2-\delta_{n,0})}{\cal N}} \left(\begin{array}{c}
  -\hbar\Omega\sqrt{n}\phi_{n-1}(\xi) \\ (E-g)\phi_n(\xi) \end{array}\right) \equiv
  \left(\begin{array}{c} \psi_+^u \\ \psi_+^l \end{array}\right), \\
 \label{phiMin}
 \psi_{nk_x-}({\bm \rho}) &=& \frac{e^{ik_xx}}{\sqrt{\pi(2-\delta_{n,0})}{\cal N}} \left(\begin{array}{c}
  (E-g) \phi_{n-1}(\xi) \\ \hbar\Omega\sqrt{n}\phi_n(\xi) \end{array}\right) \equiv
  \left(\begin{array}{c} \psi_-^u \\ \psi_-^l \end{array}\right),
\end{eqnarray}
where~$\eta=\pm 1$ labels the energy band,~$n=0,1,\ldots$ is the Landau
number,~${\cal N}=\sqrt{2E(E-g)}$,~${\bm \rho}=(x,y)$,
$\phi_n(\xi) = {\sqrt{L}C_n}{\rm H}_{n}(\xi)e^{-1/2\xi^2}$, where~${\rm H}_{n}(\xi)$ are the
Hermite polynomials, and~$C_n=1/\sqrt{2^n n!\sqrt{\pi}}$.
Energies and wave functions for~$\tau=-1$ are given below.

For~$\tau=+1$ and both~$s_z$ orientations the stationary Green's function of
the Hamiltonian in~(\ref{hH1}) is the following~$2\times 2$ matrix
\begin{equation}\label{G22}
 \hG({\bm \rho},{\bm \rho}',\cZ) = \sum_{n,k_x,\eta}
    \frac{\psi_{nk_x\eta}\psi_{nk_x\eta}^{\dagger}}{E_{nk_x\eta}-\cZ}
    \equiv \left(\begin{array}{cc} \hG^{uu} & \hG^{ul} \\ \hG^{lu} & \hG^{ll} \end{array}\right).
\end{equation}
For~$\hG^{uu}(\cZ)$ the summation over~$\eta=\pm 1$ gives
\begin{equation}
\label{Guu}
 \hG^{uu}(\cZ)=\sum_{n,k_x}\left( \frac{\psi_+^u\psi_+^{u*}}{E-\cZ'} +
      \frac{\psi_-^u\psi_-^{u*}}{-E-\cZ'}\right),
\end{equation}
in which~$\cZ' = \cZ- E_{s_z}$,~$E$ is given in~(\ref{Enkx}),
and~$\psi_{\pm}^u$,~$\psi_{\pm}^l$ are defined in~(\ref{phiPls})--(\ref{phiMin}).
The elements~$\hG^{ul}$,~$\hG^{lu}$,~$\hG^{ll}$ are obtained in an analogous way.
Calculating the sum in~(\ref{Guu}) we have
\begin{eqnarray}
\label{Guu1}
 G^{uu}({\bm \rho},{\bm \rho}', \cZ) = (\cZ' + g) \sum_{n}\!\int_{-\infty}^{\infty}\!\!
 \frac{e^{ik_x(x-x')}\phi_{n-1}(\xi) \phi_{n-1}(\xi')^*}
      {\pi(2-\delta_{n,0})(\hbar^2\Omega^2 n+g^2 - \cZ^{'2})} dk_x, \ \ \ \ \\
 \label{Gll1}
 G^{ll}({\bm \rho},{\bm \rho}', \cZ) = (\cZ' + g) \sum_{n}\!\int_{-\infty}^{\infty}\!\!
 \frac{e^{ik_x(x-x')}\phi_{n}(\xi) \phi_{n}(\xi')^*}
      {\pi(2-\delta_{n,0})(\hbar^2\Omega^2 n+g^2 - \cZ^{'2})} dk_x, \ \ \\
\label{Gul1}
 G^{ul}({\bm \rho},{\bm \rho}', \cZ) = -\sum_{n}\!\int_{-\infty}^{\infty}\!\!
 \left(\hbar\Omega \sqrt{n} \right) \frac{e^{ik_x(x-x')}\phi_{n-1}(\xi) \phi_{n}(\xi')^*}
 {2\pi(\hbar^2 \Omega^2 n + g^2 -\cZ^{'2})} dk_x.
\end{eqnarray}
To integrate over~$k_x$ we use the identity, see~\cite{GradshteinBook}
\begin{equation} \label{eHmHn}
 \int_{-\infty}^{\infty}\!\!\!\! e^{-x^2}{\rm H}_m(x+y) {\rm H}_n(x+z)dx =
      2^n\sqrt{\pi}m!z^{n'}{\rm L}_m^{n'}(-2yz),
\end{equation}
where~$n'=n-m$,~$m\leq n$, and~$L_n^{\alpha}(t)$ are the associated Laguerre polynomials.
Performing the integration over~$k_x$ we obtain
\begin{eqnarray}
\label{Guu2}
 \hG^{uu}({\bm \rho},{\bm \rho}', \cZ) = \frac{(\bZ +\bg) e^{-r^2/2+i\chi}}{2\pi\hbar \Omega L^2}
 \sum_{n=0}^{\infty} \frac{L_n^0(r^2)}{ n+1 + \bg^2-\bZ^2},\\
 \hG^{ll}({\bm \rho},{\bm \rho}', \cZ) = \frac{(\bZ+\bg) e^{-r^2/2+i\chi}}{2\pi\hbar \Omega L^2}
 \sum_{n=0}^{\infty} \frac{L_n^0(r^2)}{n + \bg^2-\bZ^2},\\
\label{Gul2}
\hG^{ul}({\bm \rho},{\bm \rho}', \cZ) =
 r_{cd}\frac{e^{-r^2/2+i\chi}}{2\pi\hbar \Omega L^2}
 \sum_{n=1}^{\infty} \frac{L_{n-1}^1(r^2)}{n+ \bg^2 - \bZ^2}, \ \ \
\end{eqnarray}
where~$\bZ=\cZ'/(\hbar\Omega)$,~$\bg=g/(\hbar\Omega)$,~$r^2 = ({\bm \rho} - {\bm \rho}')^2/(2L^2)$
and~$\chi=(x-x')(y+y')/2L^2$ is the gauge-dependent phase factor. In~(\ref{Gul2}) we
defined~$r_{ul}= [(y-y')+i(x-x')]/(\sqrt{2}L)$. For~$\hG^{lu}$ one obtains expression analogous to
that for~$\hG^{ul}$ in~(\ref{Gul2}), but with~$r_{ul}$ replaced by~$r_{lu} = [(y'-y)+i(x-x')]/(L\sqrt{2})$.

In~(\ref{Guu2})--(\ref{Gul2}) the summation over levels~$n$ is performed with the use of
formulas~6.12.4 and~6.9.4 in~\cite{ErdelyiBook}
\begin{eqnarray} \label{FPsi}
 t^{-\beta} \sum_{n=0}^{\infty} \frac{L_n^{-\beta}(t)}{n+a-\beta} &=& \Gamma(a-\beta)\Psi(a,\beta+1;t), \\
 \label{FPsiW}
 & =& \Gamma(a-\beta) e^{t/2}t^{-1/2-\beta/2}W_{\beta/2+1/2-a,\beta/2}(t),
\end{eqnarray}
where~$\Psi(a,c;t)$ is the second solution of the confluent hypergeometric equation~\cite{ErdelyiBook},
and~$W_{\kappa,\mu}(t)$ is the Whittaker function.
The series in~(\ref{FPsi}) converges for~$t>0$ and~$\beta>-1/2$.
On combining equations~(\ref{Guu2})--(\ref{FPsiW}) we set~$t=r^2$ and,
for~$\hG^{uu}$ or~$\hG^{ll}$, we take~$\beta=0$
and~$a=1+\bg^2- \bZ^2$ or~$a=\bg^2- \bZ^2$, respectively. Then we find
\begin{eqnarray}
 \label{Guu3}
 \hG^{uu}({\bm \rho},{\bm \rho}', \cZ) &=& \frac{(\bZ+g) e^{i\chi}}{2\pi\hbar\Omega L^2|r|}
          \Gamma(-\bZ^2+1+\bg^2) W_{\bZ^2-\bg^2-\frac{1}{2},0}(r^2),\ \ \ \ \\
 \label{Gll3}
 \hG^{ll}({\bm \rho},{\bm \rho}', \cZ) &=& \frac{(\cZ+g) e^{i\chi}}{2\pi\hbar\Omega L^2|r|}
        \Gamma(-\bZ^2+\bg^2) W_{\bZ^2-\bg^2 + \frac{1}{2},0}(r^2).
\end{eqnarray}
For~$\hG^{ul}$ there is~$\beta=-1$, which is beyond the convergence range of the series in~(\ref{FPsi}).
However,~$\hG^{ul}$ can be expressed as a
combination of~$\hG^{uu}$ and~$\hG^{ll}$ functions by using the
identity~$L_{n-1}^1(r^2)=n[(L_{n-1}^0(r^2)-L_{n}^0(r^2)]/r^2$, see~\cite{GradshteinBook}, which gives
\begin{eqnarray} \label{Gul3}
\hG^{ul}({\bm \rho},{\bm \rho}', \cZ) = \frac{r_{ul}}{r^2}\left(\bZ-\bg \right) (\hG^{uu} - \hG^{ll}),
\end{eqnarray}
and~$r_{ul}$ is defined in~(\ref{Gul2}). Since the Green's function is a hermitian operator,
there is~$\hG({\bm \rho},{\bm \rho}',E) = \langle {\bm \rho}|(\hH-E)^{-1}|{\bm \rho}'\rangle=
\hG({\bm \rho}',{\bm \rho},E)^{\dagger}$.
By interchanging~${\bm \rho}$ with~${\bm \rho}'$ in~(\ref{Guu3})--(\ref{Gul3}) one
finds:~$\hG^{cc}({\bm \rho},{\bm \rho}') = \hG^{cc}({\bm \rho}',{\bm \rho})^*$ with~$c\in\{u,l\}$,
and~$\hG^{ul}({\bm \rho},{\bm \rho}') = \hG^{lu}({\bm \rho}',{\bm \rho})^*$.
This verifies the hermiticity of the Green's function in~(\ref{Guu3})--(\ref{Gul3}).

Expressing the Green's function in terms of the Whittaker functions~$W_{\kappa,0}(z)$ is
very useful because the latter can be conveniently computed from the formula
\begin{eqnarray} \label{W_kappa_0}
W_{\kappa,0}(z) &=& \frac{\sqrt{z}\ e^{-z/2}}{\Gamma(1/2-\kappa)^2}
 \sum_{k=0}^{\infty}\frac{\Gamma(k-\kappa+1/2)}{(k!)^2}z^k \times \nonumber \\
 && \left[2\psi(k+1)-\psi(k-\kappa+1/2) -\ln(z)\right],
\end{eqnarray}
where~$\psi(z)=d\ln[\Gamma(z)]/dz$ and~$|{\rm arg}(z)| < 3\pi/2$, see~\cite{GradshteinBook}.
This expansion can be obtained from the Barnes integral representation
of the Whittaker function~$W_{\kappa,0}(z)$ through the calculation of residues, see e.g.~\cite{WangBook}.
Other convenient ways to calculate
the Whittaker functions are: expansion of~$W_{\kappa,0}(x)$ in terms of the Bessel
functions~\cite{AbramowitzBook}, combinations of power-series expansions for small-$x$ and
and large-$x$ approximations (see~\cite{Dodonov1975,GradshteinBook}), or numerical
solutions of the Whittaker equation with procedures given for example in Mathematica.

For large energies~$\bZ^2 > \bg^2$, the Whittaker functions in~(\ref{Guu3})--(\ref{Gul3})
have oscillating character with slowly increasing envelope.
Taking~$\kappa = \bZ^2 - \bg^2 \pm 1/2 \gg 1$ one has~\cite{dlmf}
\begin{eqnarray}
W_{\kappa,0}(x) &\simeq
& \sqrt{x}\Gamma(\kappa+1/2) \left[ \sin(\kappa \pi)J_0(2\sqrt{\kappa x}) \right. \nonumber \\
& -& \left. \cos(\kappa \pi)Y_0(2\sqrt{\kappa x}) +{\rm Env}_0(2\sqrt{\kappa x}) \right ], \label{G_pls}
\end{eqnarray}
where~$J_0(z)$,~$Y_0(z)$ are the Bessel functions in the standard notation
and~${\rm Env}_0(z) = \sqrt{J_0(z)^2+Y_0(z)^2}$.
For large~$x$ the Bessel functions in~(\ref{G_min}) decay as~$x^{-1/4}$, so that in this limit
the Green's functions in~(\ref{Guu3})--(\ref{Gll3}) diminish as~$|r|^{-1/2}$.
For~$\kappa \gg 1$ there is~\cite{GradshteinBook}
\begin{equation} \label{G_min}
 W_{-\kappa,0}(x) \simeq \left(\frac{z}{4\kappa}\right)^{1/4} e^{\kappa - \kappa \ln(\kappa)}
 \e^{-2\sqrt{\kappa x}}.
\end{equation}
For~$\bZ^2 < \bg^2$, i.e. for energies within the energy gap, the Green's functions
decay exponentially with~$|r|$.

For the~${\bm K'}$ point of the BZ the Hamiltonian~$\hH'$ reads
\begin{equation} \label{hHp}
\hH' = \left(\begin{array}{cc} g' & \hbar \Omega \hat{a}^+ \\
\hbar \Omega \hat{a} & -g' \end{array}\right) + E_{s_z}',
\end{equation}
where~$g' = \Delta/2 - s_z\tau\lambda /2$,~$E_s' = s_z\tau\lambda/2$ and~$\tau=-1$.
For given~$s_z$ there is~$g \neq g'$ because of different values of~$\tau$ in
the two valleys. Note that~$\hH'= \sigma_z\hH^{T}\sigma_z$, see~(\ref{hH3}).
The eigen-energies and eigenstates of~$\hH'$ are similar to those for~$\hH$,
[see~(\ref{Enkx})--(\ref{phiMin})]
\begin{eqnarray}
E_{nk_x\eta s_z}' &=& \eta \sqrt{n \hbar^2 \Omega^2 + g^{'2}} + E_{s_z}' \equiv \eta E' + E_{s_z}' \\
|\psi_{nk_x+}'\rangle &=& \frac{e^{ik_xx}}{\sqrt{\pi(2-\delta_{n,0})}{\cal N}^{'}} \left(\begin{array}{c}
\hbar\Omega\sqrt{n} \phi_n(\xi) \\ (E'-g') \phi_{n-1}(\xi) \end{array}\right), \\
|\psi_{nk_x-}'\rangle &=& \frac{e^{ik_xx}}{\sqrt{\pi(2-\delta_{n,0})}{\cal N}^{'}} \left(\begin{array}{c}
 (E'-g')\phi_{n}(\xi) \\ -\hbar\Omega\sqrt{n} \phi_{n-1}(\xi) \end{array}\right),
\end{eqnarray}
where~${\cal N}'=\sqrt{2E'(E'-g')}$. The stationary Green's function of the Hamiltonian~$\hH'$
is a~$2\times 2$ matrix, cf.~(\ref{G22})
\begin{equation}\label{G22p}
\hG'({\bm \rho},{\bm \rho}',\cZ) = \left(\begin{array}{cc} \hG^{ll} & -\hG^{lu} \\
 -\hG^{ul} & \hG^{uu} \end{array}\right),\end{equation}
with~$\cZ'=\cZ-E_{s_z}'$. The total Green's function operator is a~$8 \times 8$ block-diagonal matrix,
whose elements are combinations of~$\hG^{uu}$,~$\hG^{ll}$ and~$\hG^{lu}$ functions,
as given in (\ref{Guu3})--(\ref{Gul3}) with suitably chosen~$g$ and~$E_s$.

\begin{figure}
\includegraphics[width=8cm,height=8cm]{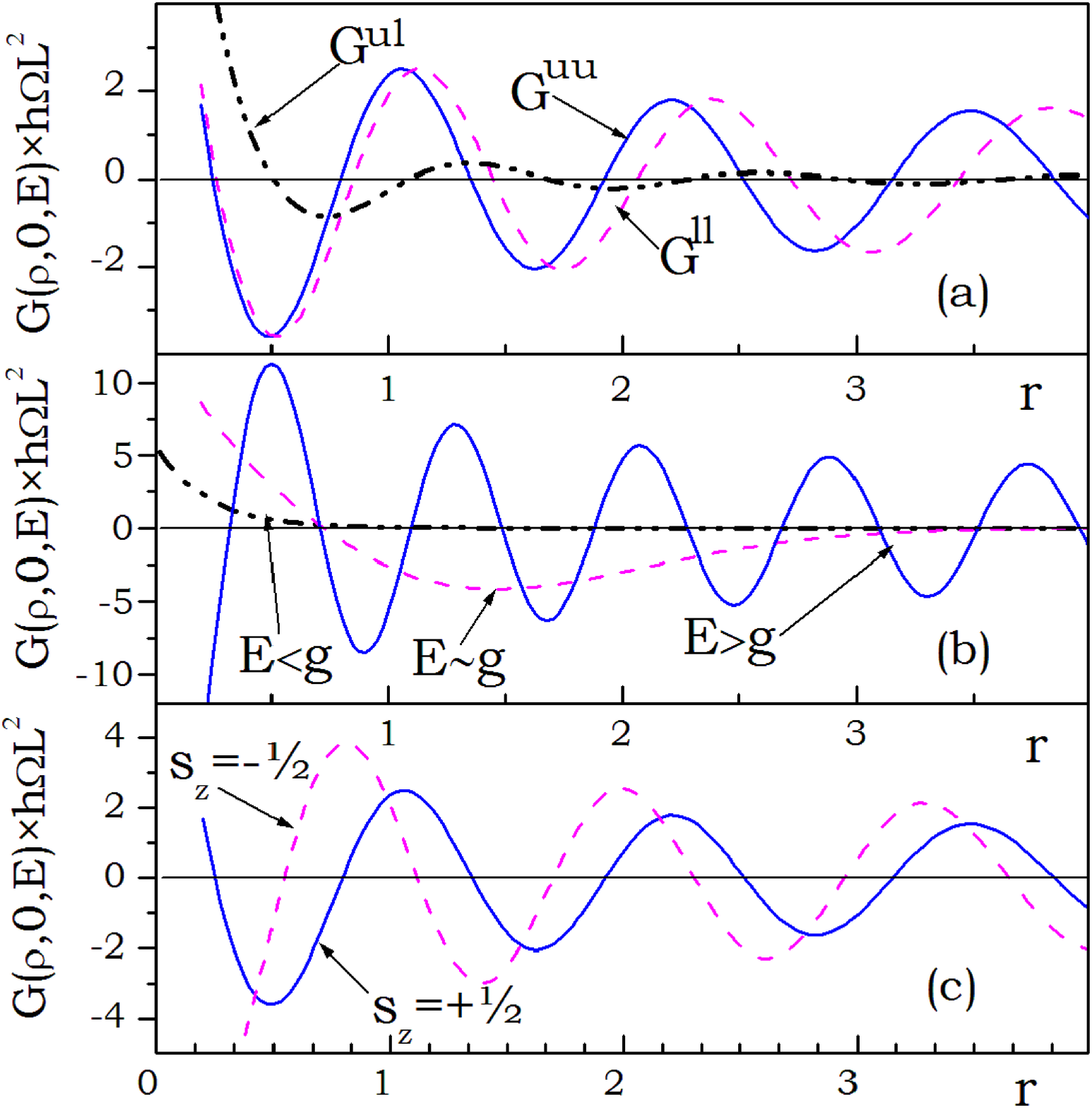}
                 \caption{Components of the gauge-independent part of
                 dimensionless Green's function~$\hbar\Omega L^2\hG({\bm \rho},0,\cZ)$ versus~$r$,
                 as given in~(\ref{Guu3})--(\ref{Gul3}) for MoS$_2$ in a magnetic field~$B=10$~T.
                 a) Elements~$\hG^{uu}$,~$\hG^{ll}$ and~$\hG^{ul}$ for~$E=0.605$~eV and~$s_z=+1/2$,
                 b) Elements~$\hG^{uu}$ for~$s_z=+1/2$.
                    Dash-dotted line:~$E=0.523$~eV~$<g$;
                    dashed line:~$E=0.555$~eV~$\simeq g$;
                    solid line:~$E=0.660$~eV. The half of energy gap is:~$g=0.531$~eV.
                 c) Elements~$\hG^{uu}$ for~$s_z=\pm 1$ for~$E=0.605$~eV
                 Distance is measured in~$r=\sqrt{x^2+y^2}/(\sqrt{2}L)$.}
\end{figure}

In Figure~1 we plot gauge-independent part of the dimensionless Green's
function~$\hbar\Omega L^2\hG^{uu}({\bm \rho},0,\cZ)$ for electrons in MoS$_2$
in a magnetic field~$B=10$~T. The parameters of the Hamiltonian (\ref{hH1}) for~MoS$_2$
are listed in Table~1. In Figure~1a we show three elements of the Green's function,
as given in~(\ref{Guu3})--(\ref{Gul3}). The diagonal elements~$\hG^{uu}$ and~$\hG^{ll}$
are similar to each other. They oscillate with the period~$\simeq L$
and gradually decrease with~$r$. The nondiagonal elements~$\hG^{ul}$
oscillate with similar period, but they decay faster
with~$r$ than~$\hG^{uu}$ or~$\hG^{ll}$, see~(\ref{Gul3}).

In Figure~1b we show~$\hG^{uu}$ for three values of~$\bZ$.
The dash-dotted line describes~$\hG^{uu}$
for energies in the energy gap. In this case~$\hG^{uu}$ decays exponentially
with~$r$ and practically disappears for~$r>L$. The dashed line corresponds to the
energies in the conduction band close to its minimum. For such energies
the Green's function slowly oscillates and gradually decays with~$|r|$,
so that only a few oscillations occur. The solid line shows~$\hG^{uu}$
for energies within the conduction band, far from its minimum.
This function oscillates and decays similarly to functions in Figure~1a.
In Figure~1c we plot~$\hG^{uu}$ for~$s_z=\pm 1$.
There is a noticeable difference between Green's functions
for different spin orientations; it becomes more pronounced for WSe$_2$ because of
much larger value of the spin-orbit splitting energy~$\lambda$, see Table~1.

We have compared numerically~(\ref{Guu3})--(\ref{Gul3}) with~(\ref{Guu2})--(\ref{Gul2})
for many randomly chosen values of~$0 < \bZ <3$ and~$0 < r< 3$,
for which~$W_{\kappa0}(z)$ was calculated using expansion~(\ref{W_kappa_0}).
After truncating the summations at~$n=1\times 10^6$ terms~(!)
we obtained only~$4$ to~$6$ significant digits of the exact results
given in terms of the Whittaker functions.
We conclude that the expansion of the Green's function in terms of the Laguerre
polynomials converges quite slowly both for zero-gap and finite-gap materials~\cite{Rusin2011}.
For large energies in the conduction band or energies in the gap, the Green's functions
in Figure~1 can be described by the asymptotic expansions of the Whittaker
functions given in~(\ref{G_pls}) and~(\ref{G_min}).

\section{Local density of states and Friedel-like oscillations}
\begin{figure}
\includegraphics[width=8cm,height=8cm]{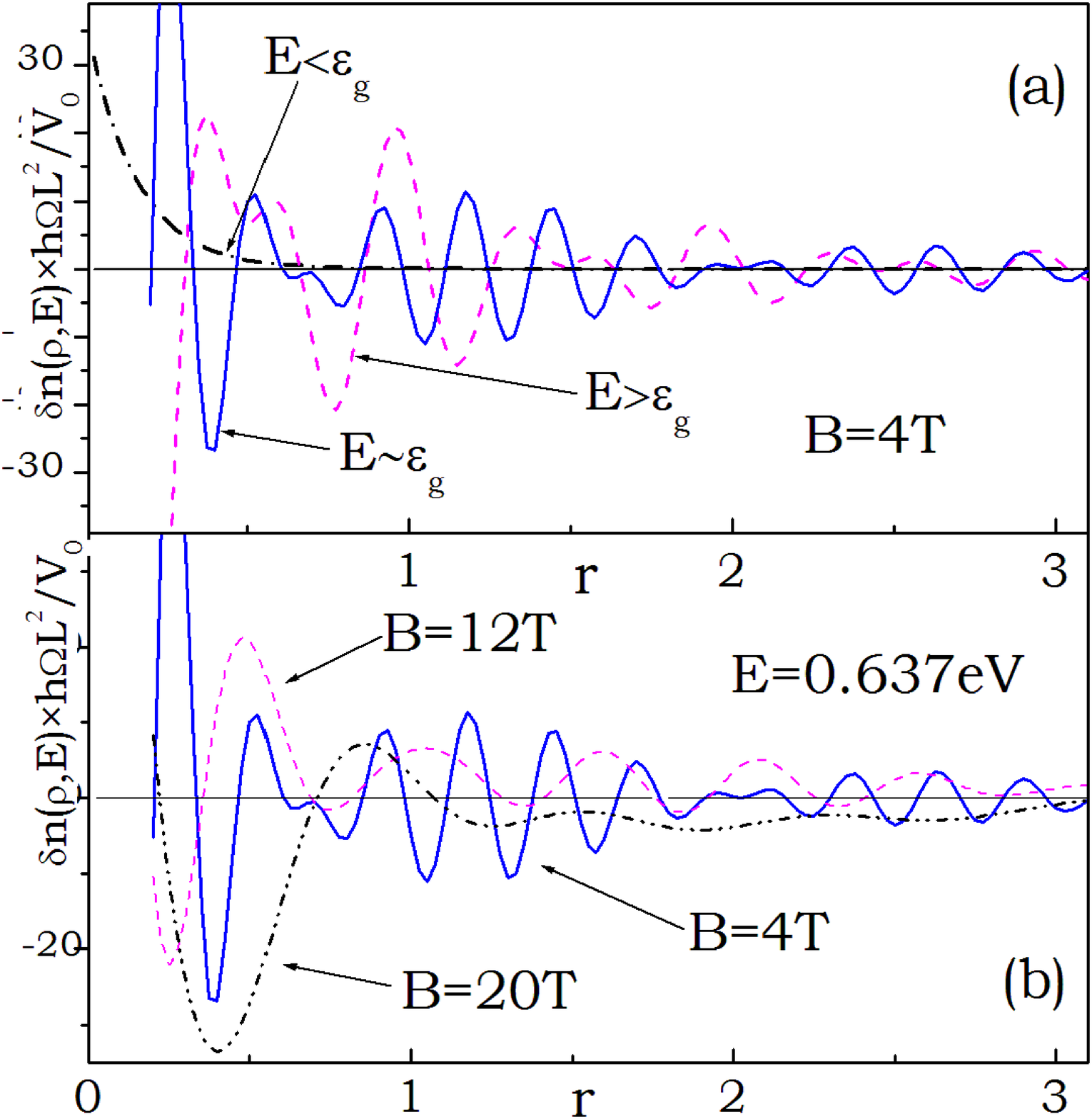}
                 \caption{Dimensionless local density of states~$\delta n(\bm \rho,\cZ)(2\pi\hbar\Omega L^2/ V_0)$
                 in MoS$_2$, as given in~(\ref{delta_n2}).
                 The average half of energy gap for two spin orientations is~$\epsilon_{g}=0.550$~eV.
                 a) Fixed magnetic field and three energy values:
                 dash-dotted line:~$E=0.505$~eV (inside the energy gap),
                 dashed line:~$E=0.584$~eV (near bottom of conduction band),
                 solid line:~$E=0.637$~eV (inside conduction band, far from minimum).
                 b) Energy~$E=0.637$~eV for three values of magnetic field.
                 Distance is measured in~$r=\sqrt{x^2+y^2}/(\sqrt{2}L)$.}
\end{figure}

\begin{figure}
\includegraphics[width=8cm,height=8cm]{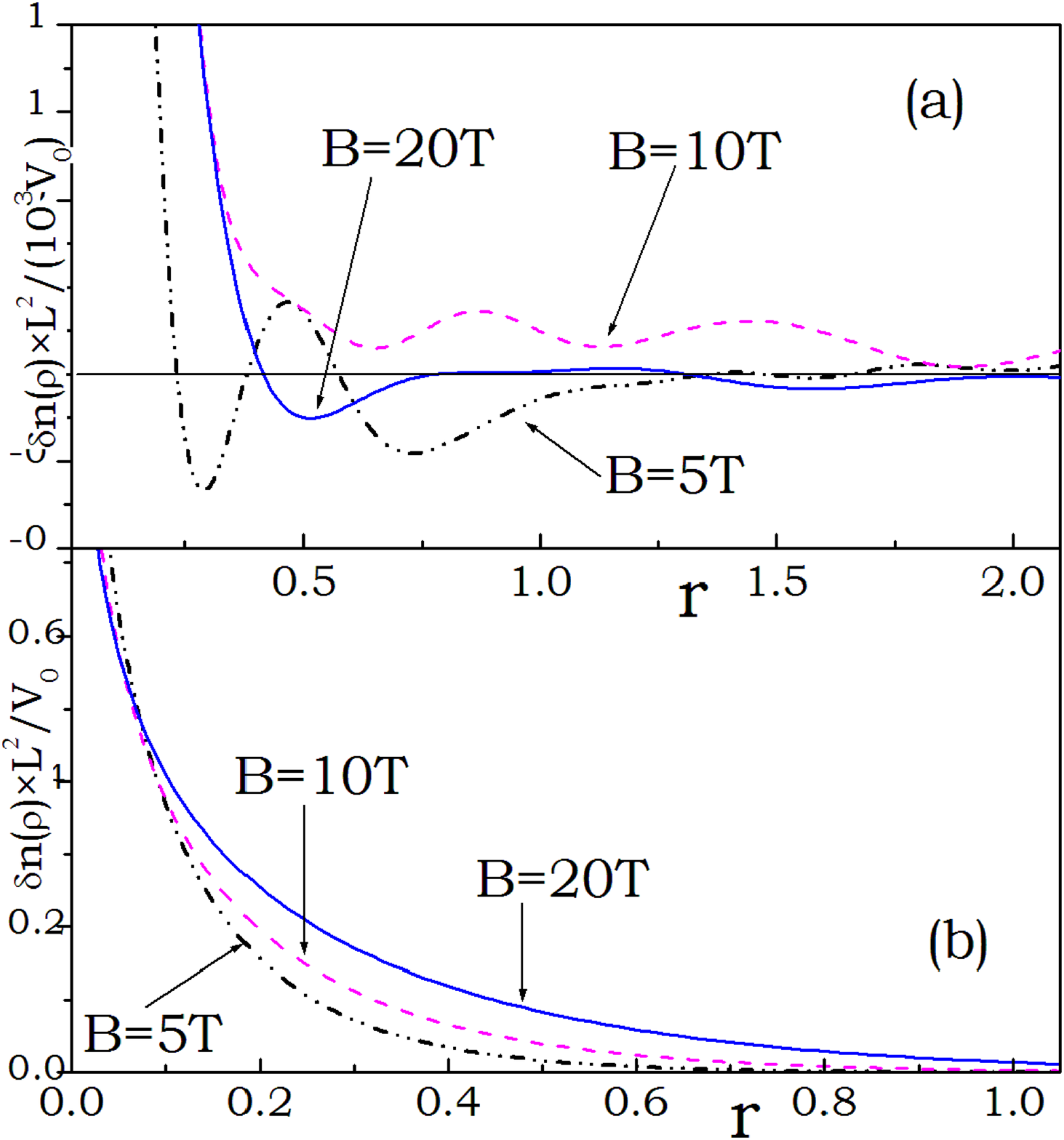}
                 \caption{Dimensionless spatial electron
                 density~$\delta n(\bm \rho)(2\pi L^2/ V_0)$ induced by the
                 delta-like neutral impurity in MoS$_2$, as given in~(\ref{delta_n3})
                 for three values of a magnetic field.
                 a) Friedel oscillations of spatial electron density.
                 Fermi level is~$0.1$~eV above the bottom of the conduction band, which
                 corresponds to~$n_e\simeq 7\times$10$^2$~cm$^{-2}$.
                 b) Exponential decay of spatial electron density.
                 Fermi level~$0.05$~eV below the bottom of the conduction band.
                 Distance is measured in~$r=\sqrt{x^2+y^2}/(\sqrt{2}L)$.}
\end{figure}

As an example of applications of the Green's function calculated in the previous section
we consider a delta-like potential of a neutral impurity localized
in the center of coordinates:~$V({\bm \rho}) = V_0\delta({\bm \rho})$~\cite{ZimanBook}.
In the presence of a magnetic field the total Green's function~$\hGG$
of electron and the impurity is obtained from the Dyson
equation:~$\hGG = \hG + \hG V \hGG$, where~$\hG$ is given in~(\ref{Guu3})--(\ref{Gul3}).
In the lowest order of the Born approximation we have
\begin{eqnarray}
 \hGG({\bm \rho}, {\bm \rho}',\cZ) &\simeq& \hG({\bm \rho}, {\bm \rho}',\cZ) +
              \int \hG({\bm \rho}, {\bm \rho}'') V({\bm \rho}'') \hG({\bm \rho}'', {\bm \rho}')
              d^2{\bm \rho}'' + \ldots \nonumber \\
              &\simeq& \hG_{{\bm \rho}, {\bm \rho}'} +
              V_0 \hG_{{\bm \rho}, {\bm 0}} \hG_{{\bm 0},{\bm \rho}'} + \ldots,
\end{eqnarray}
using the notation~$\hG_{{\bm\rho}_1,{\bm\rho}_2} = \hG({\bm\rho}_1,{\bm\rho}_2, \cZ)$.
A local density of states (LDOS) induced by the presence of impurity is
\begin{equation} \label{delta_n1}
 \delta n({\bm \rho},\cZ) = -\pi V_0 \sum_{s_z=\pm 1}
 {\rm Im} \Tr\{\hG_{{\bm \rho}, {\bm 0}}\hG_{{\bm 0},{\bm \rho}} \},
\end{equation}
assuming that~${\rm Im}\cZ > 0$. In~(\ref{G22}) and the following formulas
we set formally~$\cZ = E + i\Gamma$, so that~$\hG$ describes the causal
Green's function~\cite{ZimanBook}. Note that the unperturbed LDOS~$\propto \hG({\bm\rho}, {\bm\rho}, \cZ)$ diverges
for~$\Gamma=0$, see~(\ref{Guu2})--(\ref{Gul2}). Calculating the trace in~(\ref{delta_n1}) we find
\begin{eqnarray} \label{delta_n2}
 \delta n(\bm \rho,\cZ) &=& -\pi V_0 \sum_{\tau, s_z=\pm 1} {\rm Im}
  \left\{ \hG^{uu}_{{\bm\rho}, {\bm 0}}\hG^{uu}_{{\bm 0},{\bm \rho}} +
        \hG^{ll}_{{\bm\rho},{\bm 0}}\hG^{ll}_{{\bm 0},{\bm \rho}} \right. \nonumber \\
  &+&\left. \tau^2\left[\frac{\bZ-\bg}{\rho^2} \right]^2
       (\hG^{ll}_{{\bm \rho},{\bm 0}}-\hG^{uu}_{{\bm \rho},{\bm 0}})
       (\hG^{ll}_{{\bm 0},{\bm \rho}}-\hG^{uu}_{{\bm 0},{\bm \rho}})
 \right\}.
\end{eqnarray}
Because~$\tau^2=1$, the summation over~$\tau$ in~(\ref{delta_n2}) introduces an additional
factor of two. The summation over~$s_z$ is performed numerically,
because the spin splitting of the valence bands modifies the energy gaps~$g$
and energies~$E_{s_z}$ for the two spin orientations, see~(\ref{hH1}).

In Figure~2 we show calculated~$\delta n({\bm \rho},\cZ)$ for several values of
magnetic field and energies~$\cZ$ assuming finite Landau level widths~$\Gamma \sim 1-2$~meV,
corresponding to typical values in quasi two-dimensional systems.
It is seen from (\ref{delta_n2}) that~$\delta n$ is linear in~$V_0$,
so we can choose~$V_0$ arbitrarily. In Figure~2a we show dimensionless induced local density of
states~$\delta n({\bm \rho},\cZ)(2\pi\hbar L^2/V_0)$
for three energy values:~$E$ in the energy
gap,~$E$ just above band edge and~$E$ inside the conduction band.
The induced LDOS resembles the behavior of Green's functions in Figure~1.
For energies inside the gap,~$\delta n$ decays exponentially with the distance~$r$.
For energies inside the conduction band~$\delta n$ oscillates and
decreases with~$r$, much faster than the functions~$\hG^{uu}$ and~$\hG^{ll}$.
The oscillation period is a fraction of magnetic length~$L$ and increases with energy.
The oscillations have irregular character and disappear for~$r>3$.

In Figure 2b we plot~$\delta n({\bm \rho},\cZ)$ for three values of magnetic field.
The periods of oscillations of the induced LDOS increase with magnetic field,
but for large fields the oscillations disappear for~$r>2$.
The irregular character of~$\delta n$ oscillations results from the presence of two different
energy gaps for opposite spin orientations, leading to oscillations with two different
frequencies accompanied by fast decay of oscillations with~$r$.
In Figure~2b the induced density of states depends on the dimensionless
parameter~$r/(\sqrt{2}L)$, which is proportional to~$\sqrt{B}$.
This complicates the dependence of~$\delta n$ on magnetic field.

Having calculated LDOS one can obtain the additional spatial density of states~$\delta n({\bm r})$ induced by
the impurity at~$T=0$
\begin{equation} \label{delta_n3}
 \delta n({\bm r}) = \int _{-\infty}^{\epsilon_F}\delta n(\bm \rho,\cZ) dE,
\end{equation}
where~$\epsilon_F$ is the Fermi energy. For given electron concentration~$n_e$,
the Fermi level in a magnetic field oscillates with~$B$, see e.g.~\cite{Kubisa2015}.
However, one can roughly estimate the position
of the Fermi level in a magnetic field by its value at $B=0$.
Assuming the Fermi level in the conduction band one has
\begin{eqnarray} \label{nepm}
 n_{es_z} = \int_{0}^{\epsilon_F} g_{es_z}(E) dE, \\
 g_{es_z}(E) = \frac{1}{(2\pi)^2}\int \delta(E-E_{k s_z}) d^2 {\bm k} ,
\end{eqnarray}
where~$E_{k s_z}$ is given in~(\ref{eH1}). Then we find
\begin{equation} \label{eF}
 \epsilon_F = E_{s_z} + \sqrt{g^2 + 4\pi a^2t^2 n_{es_z}}.
\end{equation}
For electrons at the~${\bm K}$ point of BZ there is:~$E_{s_z}= \lambda s_z$,
while at the~${\bm K'}$ point of BZ one has:~$E_{s_z}= -\lambda s_z$. Thus, by
summing contributions to electron density arising from both points of BZ
we find:~$n_{e+}=n_{e-}=n_e/2$.

In Figure~3 we plot the induced electron density~$\delta n({\bm r})$, as given
in~(\ref{delta_n3}) calculated for three values of magnetic field and~$\epsilon_F$.
The latter is measured from the bottom of the conduction band:~$E_0=\Delta/2 -\lambda/2$.
In Figure~3a the Fermi level is~$0.1$~eV above~$E_0$. The concentration~$n_{e+}$
calculated from~(\ref{eF}) is:~$n_{e+}=3.5\times$10$^{-12}$~cm$^{-2}$
and total electron concentration is~$n_e=n_{e+}+n_{e-} =7.0 \times$10$^{-12}$~cm$^{-2}$,
which is a representative value for~$n$-doped MoS$_2$. The induced electron density
oscillates with~$r$ and vanishes for~$r>2$. The oscillations in Figure~3a are analogous to
the Friedel oscillations in free electron gas~\cite{Friedel1952}.
The signs and phases of oscillations are sensitive to the magnitude of
magnetic field. Note that for~$B=10$~T the induced density oscillates between
positive values only, while for the remaining cases the induced density takes both positive
and negative values.

Comparing~(\ref{eH1}) with~(\ref{eF}) we
find:~$k_F=\sqrt{2\pi n_{e}}=0.066$~\AA$^{-1}$. The expected period of Friedel
oscillations is:~$T_F=2\pi/(2k_F)=47.4$~\AA~\cite{Friedel1952}. For~$B=5$~T the
magnetic length is:~$L=115$~\AA, so that:~$T_F=0.41\ L$, and similar value is found
in Figure~3a. Taking~$B=10$~T we obtain:~$T_F=0.58\ L$, while for~$B=20$~T
there is:~$T_F=0.83$~\AA. These results justify the interpretation of oscillations in
Figure~3 as the Friedel oscillations.

In Figure~3b we plot the induced spatial electron density for
the Fermi level~$0.05$~eV below the bottom of the conduction band.
In this case~$\delta n({\bm r})$ decays exponentially and vanishes
for distances larger than the magnetic length~$L$. The decay of
electron density is very similar for different values of a magnetic field.
The results are analogous to those obtained for an isolated
delta-like impurity in free space, in which the exponential decay of
electron density takes place~\cite{Atkinson1975}.

\section{Summary}
To summarize, we calculated the Green's functions for electrons in group-VI dichalcogenides
in the presence of a magnetic field and expressed them in terms of the Whittaker functions.
The obtained formulas allow one to compute the Green's functions using quickly
converging expansions. For energies in
the energy gap the Green's functions decay exponentially with the distance, while for
energies in the conduction band the Green's function oscillate
with a slowly vanishing envelope. The obtained Green's functions
are used for a calculation of the induced density of states for a delta-like
impurity potential in~MoS$_2$ in the presence of a magnetic field within
the lowest order of the Born series. The presence of spin-orbit splitting of the
valence band gives two values of energy gaps,
and the resulting density of states oscillates with two different frequencies.
For the Fermi level in the conduction bands the electron density induced
by the impurity experiences Friedel-like oscillations, while
for the Fermi level in the energy gap it decays exponentially with the distance
from impurity. The results obtained in this work can
be extended to other group-VI dichalcogenides.

\hspace*{1em}


\begin{thebibliography}{99}
\bibitem{Dodonov1975}     Dodonov V V, Malkin I A, and Man'ko V I 1975 Phys. Lett. A {\bf 51} 133
\bibitem{Gusynin1995}     Gusynin V P, Miransky V A and Shovkovy I A 1995 Phys. Rev. D {\bf 52} 4718
\bibitem{Gorbar2002}      Gorbar E V, Gusynin V P, Miransky V A and Shovkovy I A 2002 Phys. Rev. B {\bf 66} 045108
\bibitem{Murguia2010}     Murguia G, Raya A, Sanchez A and Reyes E 2010 Am. J. Phys. {\bf 78} 700
\bibitem{Horing2009}      Horing N J M and Liu S Y 2009 J. Phys. A {\bf 42} 225301
\bibitem{Pyatkovskii2011} Pyatkovskii P K and Gusynin V P 2011 Phys. Rev. B {\bf 83} 075422
\bibitem{Gamayun2011}     Gamayun O V, Gorbar E V and Gusynin V P 2011 Phys. Rev. B {\bf 83} 235104
\bibitem{Rusin2011}       Rusin T M and Zawadzki W 2011 J. Phys. A {\bf 44} 105201
\bibitem{Ardenghi2015}    Ardenghi J S, Bechthold P, Gonzalez E, Jasen P and Alfredo J 2015 Eur. Phys. J. B {\bf 88} 47
\bibitem{Rubio2016}       Gutiérrez-Rubio A, Stauber T, Gómez-Santos G, Asgari R and Guinea F 2016 Phys. Rev. B {\bf 93} 085133
\bibitem{Liu2013}         Liu G B, Shan W Y, Yao Y, Yao W and Xiao D 2013 Phys. Rev. B {\bf 88} 085433
\bibitem{Xiao2012}        Xiao D, Liu G B, Feng W , Xu X and Wang Y 2012 Phys. Rev. Lett. {\bf 108} 196802
\bibitem{Rensink1968}     Rensink M. E. 1968 Phys. Rev. {\bf 174} 744
\bibitem{Glasser1969}     Glasser M. L. 1969 Phys. Rev. {\bf 180} 942
\bibitem{Simion2005}      Simion G E and Giuliani G F 2005 Phys. Rev. B {\bf 72} 045127
\bibitem{GradshteinBook}  Gradshtein I S and Ryzhik I M 2007 {\it Table of Integrals, Series, and
                          Products (Ed Jeffrey A and Zwillinger D 7th edition)} (New York: Academic Press) p~804, p~1001, p~1026, p~1027
\bibitem{ErdelyiBook}     Erdelyi A (ed.) 1953 {\it Higher Transcendental Functions}, vol. I (New York: McGraw-Hill)
\bibitem{WangBook}        Wang I X and Guo D R 1989 {\it Special Functions} (World Scientific, Singapore)
\bibitem{AbramowitzBook}  Abramowitz M and Stegun I (ed) 1972 {\it Handbook of Mathematical Functions} (New York: Dover)
\bibitem{dlmf}            http://dlmf.nist.gov/13.21 2016 eq. 13.21.2
\bibitem{ZimanBook}       Ziman J M 1969 {\it Elements of Advanced Quantum Theory} (Cambridge: University Press) p.~131
\bibitem{Kubisa2015}      Zawadzki W and Lassing R 1984 Surf. Sci. {\bf 142} 225
\bibitem{Friedel1952}     Friedel J 1952 Philos. Mag. {\bf 43} 153
\bibitem{Atkinson1975}    Atkinson D A and Crater H W 1975 Am. J. Phys. {\bf 43} 301

\end{thebibliography}
\end{document}